\documentclass[10pt,conference]{IEEEtran} 
\IEEEoverridecommandlockouts
\usepackage{cite}
\usepackage{amsmath,amssymb,amsfonts}
\usepackage{graphicx}
\usepackage{textcomp}
\usepackage{subfig}
\usepackage{xcolor}
\usepackage{xspace}
\usepackage{xspace}
\usepackage{bm}
\usepackage{adjustbox}
\usepackage{tabularx}
\usepackage{color}
\usepackage{algorithm}
\usepackage{algpseudocode}
\usepackage{listings}
\usepackage{nccmath}
\usepackage{subfig}
\usepackage{comment}
\usepackage[font=small, labelfont=bf]{caption}
\usepackage{mdwtab}
\usepackage{multirow}
\usepackage{xspace}
\usepackage{graphicx}
\usepackage{multirow}
\usepackage{tabularx}
\def\BibTeX{{\rm B\kern-.05em{\sc i\kern-.025em b}\kern-.08em
    T\kern-.1667em\lower.7ex\hbox{E}\kern-.125emX}}
\begin{document}

\title{Code Librarian: A Software Package Recommendation System\\
}

\makeatletter
\DeclareRobustCommand\onedot{\futurelet\@let@token\@onedot}
\def\@onedot{\ifx\@let@token.\else.\null\fi\xspace}

\def\eg{\emph{e.g}\onedot} \def\Eg{\emph{E.g}\onedot}
\def\ie{\emph{i.e}\onedot} \def\Ie{\emph{I.e}\onedot}
\def\cf{\emph{c.f}\onedot} \def\Cf{\emph{C.f}\onedot}
\def\etc{\emph{etc}\onedot} \def\vs{\emph{vs}\onedot}
\def\wrt{w.r.t\onedot} \def\dof{d.o.f\onedot}
\def\etal{\emph{et al}\onedot}
\makeatother

\author{\IEEEauthorblockN{Lili Tao,
Alexandru-Petre Cazan, Senad Ibraimoski and
Sean Moran}
\IEEEauthorblockA{JP Morgan Chase\\
Email: \{lili.tao,alexandru-petre.cazan,senad.ibraimoski,sean.j.moran\}@jpmchase.com}}
\maketitle

\maketitle

\begin{abstract}
 The use of packaged libraries can significantly shorten the software development life cycle by improving the quality and readability of code. In this paper, we present a recommendation engine called Code Librarian for open source libraries. A candidate library package is recommended for a given context if: 1) it has been frequently used with the imported libraries in the program; 2) it has similar functionality to the imported libraries in the program; 3) it has similar functionality to the developer’s implementation, and 4) it can be used efficiently in the context of the provided code. We apply the state of the art CodeBERT-based model for analysing the context of the source code to deliver relevant library recommendations to users.
\end{abstract}


\section{Introduction}
Reusing existing software libraries brings many benefits, including the acceleration of software development and an increase in the quality and readability of code. In this paper, we introduce \emph{Code Librarian}, a software library recommendation system that uses machine learning techniques to suggest relevant open source libraries based on the context of the code already written by a developer~\cite{ouni2017search, sun2020req2lib}.
For Python developers there are more than 350,000 libraries \cite{pypi} available on PyPi and new library packages are frequently added. In addition, there is rapid evolution of standard library practices across various tasks. For example, in the field of Natural Language Processing (NLP) the commonly used libraries quickly expanded from \textit{scikit-learn} and \textit{genism} to \textit{bertopic}, \textit{top2vec}, \textit{octis}, based on recent advances in NLP. Librarian is an intelligent coding assistant that helps developers find and reuse quality code and components. 

\section{Approach and Methodology}

\begin{figure*}
\centerline{\includegraphics[scale=0.30]{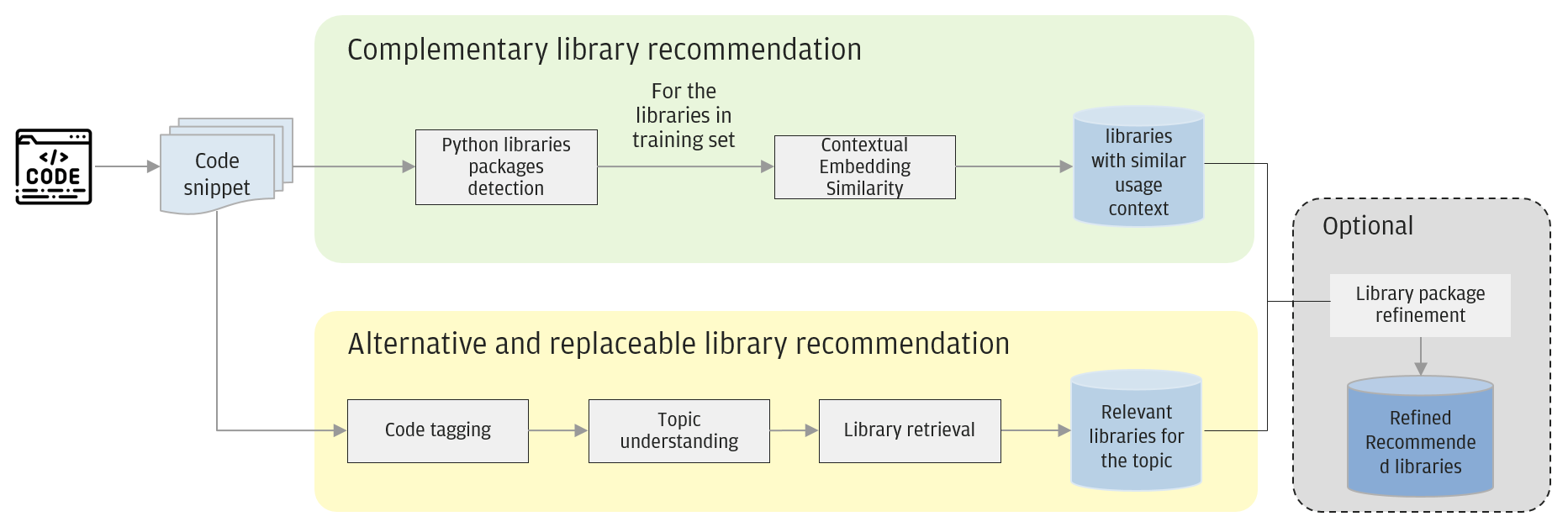}}
\caption{Librarian takes a Python script or Notebook, analyses the code and recommends software libraries.}
\label{fig:overview}
\end{figure*}

Figure \ref{fig:overview} shows the approach: a) recommendation of \emph{complementary libraries} by learning which libraries are used most frequently with those imported; b) recommendation of \emph{replaceable libraries} that can replace functionally similar code.\subsection{Complementary library recommendation}
 \noindent\textbf{Learning embeddings for library packages:} To discover complimentary libraries we learnt a contextual embedding of libraries based on their co-occurrence in the same scripts. We followed \cite{theeten2019import2vec} for learning the vector representation of library packages in which a skip-gram model \cite{mikolov2013distributed} is used to learn embeddings for libraries based on their usage context. A pair of imported libraries are deemed a positive example when the target library co-occurred with the context library within a file of at least one project. A negative pair are libraries that were rarely imported together in any source file of any project in the dataset. Cosine similarity between the embeddings is used to find very similar, and therefore, complimentary libraries. 
 
 \textbf{Out of sample extension:} For new library packages not included in the training data, rather than re-train the model, an embedding of the new package is learnt by projecting it into the latent space. The new embedding can be calculated by the weighted average of $N$ co-occurring packages in the same file, with the weight representing the number of times the pair appeared together:
$\frac{\sum_{i=1}^{N}w_iP_i}{\sum_{i=1}^{N}w_i}$, where weights $w_i$ is the number of times the unseen library co-occurred with library $P_i$

\subsection{Alternative library recommendation}
Understanding the topic and functionality of source code assists with the selection of relevant libraries. We leverage CodeBERT \cite{feng2020codebert} to learn the contextual representation of the code and capture the semantic connection between natural language and programming language.
CodeBERT is applied to generate a text description of each function or Jupyter notebook cell for IPython notebook files. We  concatenate the text descriptions and use those as a query. The query is used to retrieve matching libraries based on their description using a vector-space retrieval method (bag-of-words, TF-IDF).

\subsection{Deployment of Librarian}
\begin{figure*}
\centerline{\includegraphics[scale=0.4]{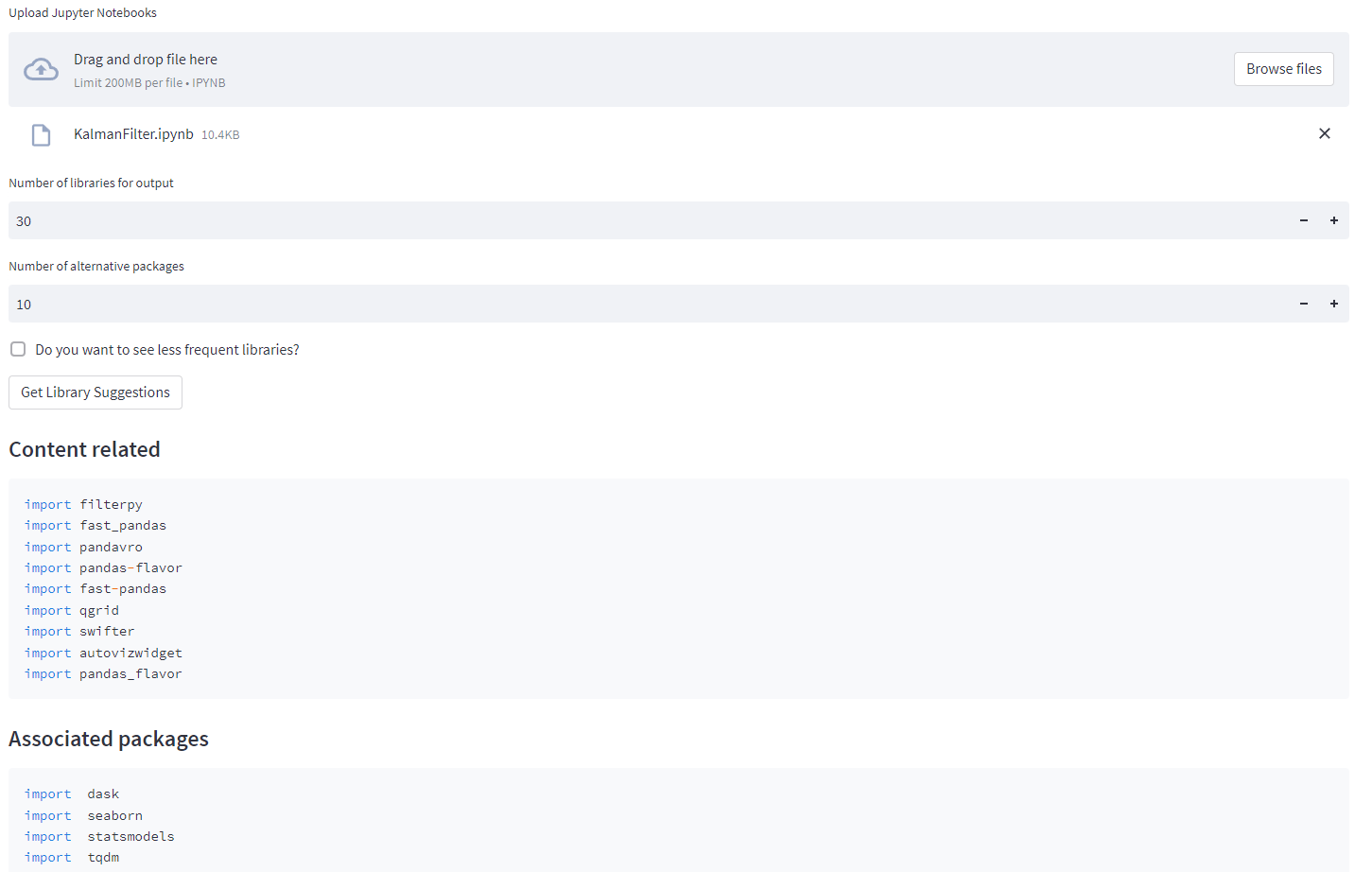}}
\caption{Streamlit demo: User uploads a notebook file leveraging streamlit frontend and obtains the library recommendations on the same page}
\label{fig:implementations1}
\end{figure*}

\begin{figure*}
\centerline{\includegraphics[scale=0.5]{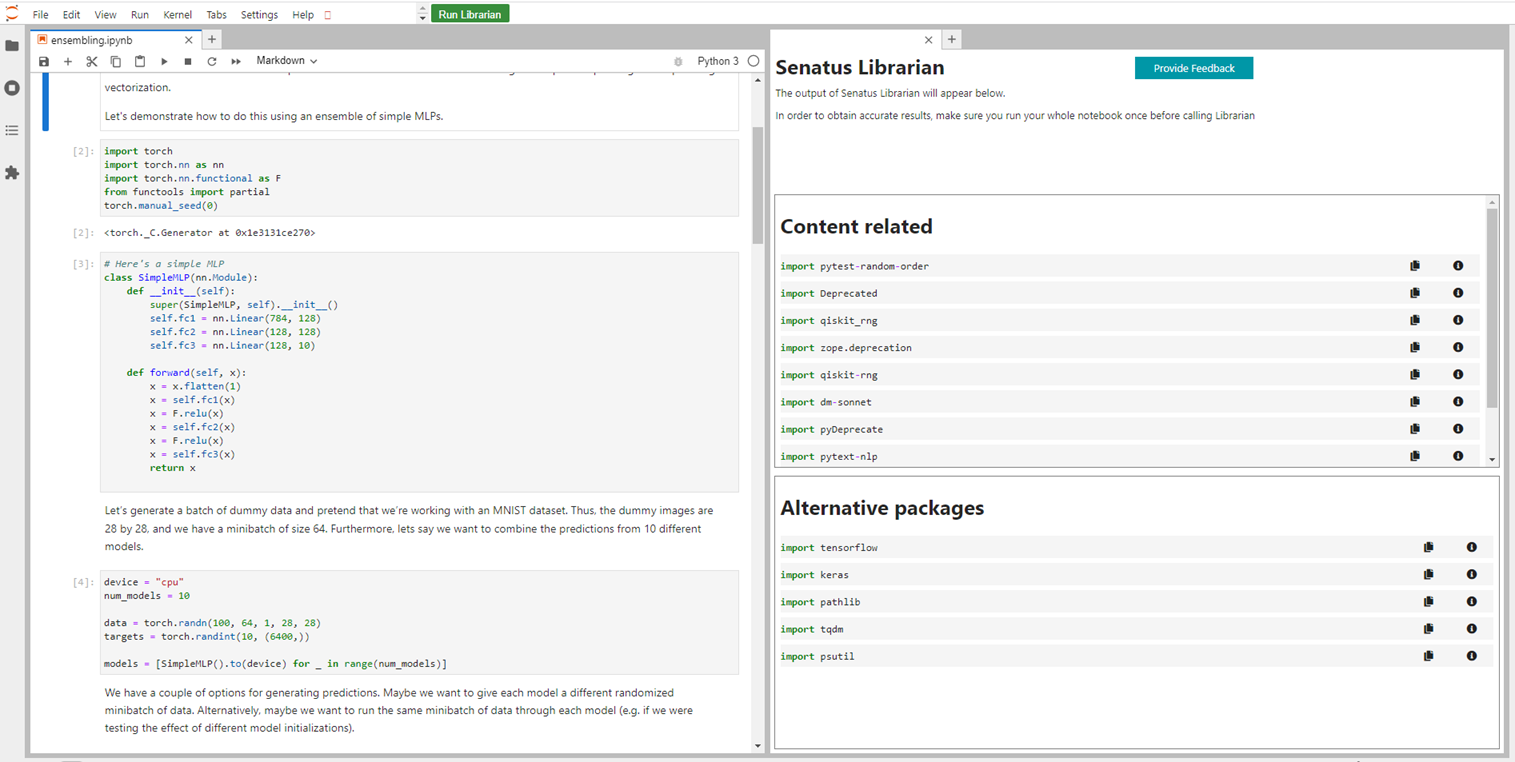}}
\caption{System listens to the cell update events and displays library recommendations in a sub-panel shown on the right side}
\label{fig:implementations2}
\end{figure*}

We  developed a demo shown in Figure \ref{fig:implementations1}. CodeBERT was packaged and deployed on AWS Sagemaker, while the main application was deployed as a service on a Kubernetes cluster. The user receives library recommendations after uploading a Jupyter notebook file. For a more seamless user experience we built a Jupyter notebook extension which subscribes to cell change events and recommends libraries in real time. On every cell update event, the current notebook sourcecode is sent to a CodeBERT model for inference and the results are shown in a sub-panel (Figure \ref{fig:implementations2}).

\section{Experimental Results}
The proposed system has been evaluated on 375,128 publicly available Python files from GitHub, and on 11,893 files from a proprietary repository. \vspace{0.1in}

\noindent\textbf{Complementary library recommendation:} to evaluate complementary library recommendation, we randomly remove one of the imported libraries from the Python file, and use the remainder to predict the removed library. Top-5 and top-10 accuracy are reported. As summarised in Table \ref{tab:complementary}, similar performance is achieved for both public available repository and the proprietary repository. 
\vspace{0.1in}

\noindent\textbf{Alternative library recommendation:} To evaluate alternative library recommendation, we create two types of truth labels. For internal data, \textit{hard labels} are created by manually annotating the expected libraries based on the functionality of the code. The ground truth libraries were annotated by a group of data scientists across 97 python scripts. Due to the manually intensive labelling process, we introduce \textit{soft labels} for which the docstrings in the file are used as ground truth description. The same information retrieval algorithm is applied to obtain a list of ground truth libraries for evaluation. Both hard and soft labels are applied to internal repositories, but only soft labels are applied to public repositories. Results are shown in Table \ref{tab:alternative}. 
\vspace{0.1in}

\noindent\textbf{User Evaluation:} A trial was conducted with a team of data scientists, all permanent employees of JP Morgan Chase who had on average 2.5 years of experience. For each recommended library a user gives feedback by choosing from one of three options; 1) Yes: the recommendation is useful and you are willing to use the recommended library, 2) Relevant but not required: the recommendation is relevant, but you might not want to switch to it; 3) Not relevant: recommendation is not relevant. Table \ref{tab:trial} shows the percentage of each options in the internal trial. Overall an encouraging 31\% of users reported the recommended libraries are “useful and willing to use”, 51\% reported as “relevant but may not want to switch to”, and 11\% reported as “not relevant”.

\begin{table}[]
\begin{centering}
\begin{tabular}{l|l|l} \hline
       & Public & Internal \\ \hline
Top-5  & 47.02  & 42.96    \\ 
Top-10 & 60.67  & 59.88    \\ \hline
\end{tabular}
\caption{Complementary library prediction results.}
\label{tab:complementary}
\end{centering}
\end{table}

\begin{table}[]
\begin{centering}
\begin{tabular}{l|l|l} \hline
           & Public & Internal \\ \hline
Hard label & /      & 30.77    \\ 
Soft label & 63.70  & 46.15    \\ \hline
\end{tabular}
\caption{Alternative library prediction results.}
\label{tab:alternative}
\end{centering}
\end{table}

\begin{table}[]
\centering
\begin{tabular}{lll|lll} \hline
\multicolumn{3}{l|}{Complementary library}                                                & \multicolumn{3}{l}{Alternative library}                                                  \\ \hline
\multicolumn{1}{l|}{Yes}  & \multicolumn{1}{l|}{Relevant} & Not relevant & \multicolumn{1}{l|}{Yes}  & \multicolumn{1}{l|}{Relevant} & Not relevant \\ \hline
\multicolumn{1}{l|}{44\%} & \multicolumn{1}{l|}{41\%}                      & 12\%         & \multicolumn{1}{l|}{27\%} & \multicolumn{1}{l|}{57\%}                      & 11\%         \\ \hline
\end{tabular}
\caption{JP Morgan Chase internal trial results}
\label{tab:trial}
\vspace{-6mm}
\end{table}

\section{Threats to Validity}\label{sec:threats}
The user study focused on one particular team within JP Morgan Chase. While a good indicator of the usefulness of Librarian, we would suggest an expansion of the study to more groups. Additionally the library recommendation should take into account other factors such as vulnerabilities, overhead, abilities~\etc,  which were not considered here. Finally, the study has initially focused on Python which is popular with data scientists, with future work expanding application to other languages.

\section{Conclusion}
We present a recommender system (``Code Librarian'') that suggests relevant libraries based on the code that a developer has written. Our results demonstrate the potential of Librarian in accelerating software development and encouraging code reuse. Our study opens up many avenues for future work, and we intend to begin by tackling the tasks outlined in Section~\ref{sec:threats}. 

\bibliographystyle{IEEEtran}
\bibliography{sample}
\end{document}